\documentclass[manuscript, nonacm]{acmart}
\AtBeginDocument{%
  }

\setcopyright{acmlicensed}
\copyrightyear{2025}
\acmYear{2025}
\acmDOI{XXXXXXX.XXXXXXX}
\acmConference[OzCHI '25]{the 37th Australian Conference on Human-Computer Interaction}{2025}{Sydney, Australia}
\acmISBN{978-1-4503-XXXX-X/2025/12} 

\usepackage{tikz}
\usetikzlibrary{positioning, arrows.meta, shapes.geometric}

\begin{document}
\captionsetup[table]{font=normalsize, labelfont=bf}
\captionsetup[figure]{font=normalsize, labelfont=bf}

\thanks{This is the author-prepared version of a paper accepted to OZCHI 2025. The official version will be published by ACM.}

\begin{center}
\fbox{
\parbox{0.96\linewidth}{
\textbf{Please cite as follows:}  
Mahmud, R., Berkovsky, S., Prasad, M., and Kocaballi, A. B. (2025). \textit{Understanding User Preferences for Interaction Styles in Conversational Recommender System: The Predictive Role of System Qualities, User Experience, and Traits}. In \textit{Proceedings of the Australian Human–Computer Interaction Conference (OZCHI 2025)}. ACM.
}
}
\end{center}
\vspace{1em}

\title [Predicting Interaction Preference in Conversational Recommender] {Understanding User Preferences for Interaction Styles in Conversational Recommender System: The Predictive Role of System Qualities, User Experience, and Traits}
\author{Raj Mahmud}
\email{raj.mahmud@uts.edu.au}
\affiliation{%
  \institution{School of Computer Science, University of Technology Sydney}
  \city{Sydney}
  \state{NSW}
  \country{Australia}
}

\author{Shlomo Berkovsky}
\affiliation{%
 \institution{Australian Institute of Health Innovation, Macquarie University}
 \city{Sydney}
 \state{NSW}
 \country{Australia}}

\author{Mukesh Prasad}
\affiliation{%
  \institution{School of Computer Science, University of Technology Sydney}
  \city{Sydney}
  \state{NSW}
  \country{Australia}}

\author{A. Baki Kocaballi}
\email{Baki.Kocaballi@uts.edu.au}
\affiliation{%
  \institution{School of Computer Science, University of Technology Sydney}
  \city{Sydney}
  \state{NSW}
  \country{Australia}
}

\renewcommand{\shortauthors}{Mahmud et al.}

\begin{abstract}
Conversational Recommender Systems (CRSs) deliver personalised recommendations through multi-turn natural language dialogue and increasingly support both task-oriented and exploratory interactions. Yet, the factors shaping user interaction preferences remain underexplored. In this within-subjects study (\(N = 139\)), participants experienced two scripted CRS dialogues, rated their experiences, and indicated the importance of eight system qualities. Logistic regression revealed that preference for the exploratory interaction was predicted by enjoyment, usefulness, novelty, and conversational quality. Unexpectedly, perceived effectiveness was also associated with exploratory preference. Clustering uncovered five latent user profiles with distinct dialogue style preferences. Moderation analyses indicated that age, gender, and control preference significantly influenced these choices. These findings integrate affective, cognitive, and trait-level predictors into CRS user modelling and inform autonomy-sensitive, value-adaptive dialogue design. The proposed predictive and adaptive framework applies broadly to conversational AI systems seeking to align dynamically with evolving user needs.
\end{abstract}




\maketitle

\section{Introduction}
Conversational Recommender Systems (CRSs) support users in discovering and selecting items through natural-language dialogue. Conceptually, CRSs comprise three core modules \cite{jannach2021survey}. First, a dialogue interface module integrates natural language understanding (NLU) and natural language generation (NLG). The NLU component processes user utterances to extract intents, entities, and contextual cues, while NLG generates fluent, coherent responses aligned with conversation history. Second, a dialogue management (DM) module tracks the dialogue state and selects appropriate system actions, such as offering suggestions, requesting clarification, or adjusting interaction strategy. Third, a recommendation engine (RE) selects candidate items based on user preferences and context. While traditional CRSs rely on collaborative filtering or content-based retrieval, retrieval-augmented generation (RAG) systems enhance this step by integrating external knowledge bases and large language models (LLMs) \cite{xu2024leveraging}.

Early CRSs often employed rule-based dialogue policies with system-led initiative to efficiently guide users toward relevant items \cite{Christakopoulou2016, pramod2022conversational}. In parallel, \emph{exploratory} interaction modes, featuring open-ended suggestions, probing questions, and invitations to browse, emerged from research on interactive search and critiquing systems \cite{marchionini2006exploratory, rieser2011reinforcement}. More recently, LLM-powered CRSs have expanded this space by enabling dynamic elaboration, contextual questioning, and on-the-fly generation of diverse recommendations \cite{deldjoo2404review, xu2024leveraging}. These systems support both \emph{task-oriented} and \emph{exploratory} strategies, offering greater flexibility in tone, initiative management, and exploration \cite{zhao2024recommender, li2023large}. Consequently, recent studies have shifted focus from accuracy-centric optimisation to dialogue strategy design, adaptive user modelling, and UX-oriented evaluation frameworks \cite{farshidi2024understanding, jannach2023evaluating, wu2024survey, jin2024crs}. While these efforts indicate the importance of interaction quality in shaping user satisfaction, trust, and continued use, it remains unclear how users experience different interaction styles, especially in relation to affective and cognitive dimensions, and how these preferences vary across individual traits.

Although the contrast between \emph{task-oriented} and \emph{exploratory} dialogue strategies has inspired various system design approaches \cite{chen2012critiquing, mahmood2009improving}, user preferences for these modes remain poorly understood. Clarifying when and why individuals favour each style is essential for tailoring interaction strategies to user expectations. While frameworks such as ResQue \cite{pu2011user} and CRS-Que \cite{jin2024crs} effectively capture retrospective evaluations of satisfaction and trust, they do not anticipate a user’s preferred interaction style in advance. Moreover, individual factors, including control preferences, prior experience with CRSs, and demographic traits, shape users’ perceptions of the system and influence their interaction style preferences. However, their impact on strategy preference has not been examined systematically \cite{folstad2020users, tsai2021effects}.
To address these gaps, we investigate three research questions:  

\textbf{RQ1:} Which user-perceived CRS qualities and UX dimensions predict interaction preference?

\textbf{RQ2:} What user profiles emerge based on CRS qualities, and how do they differ in interaction preference?

\textbf{RQ3:} Do user characteristics moderate interaction preferences in CRS conversations?

We make four contributions to the theory and design of CRSs. First, we introduce a controlled interaction‐preference paradigm contrasting \emph{task-oriented} and \emph{exploratory} dialogue strategies, enabling systematic analysis of affective and cognitive factors. Second, we extend the CRS-Que framework \cite{jin2024crs} to predictive preference modelling by linking perceived system qualities and UX appraisals to interaction choices. Third, clustering analysis reveals five latent user profiles that can inform value-based, trait-sensitive personalisation strategies. Fourth, we identify moderation effects of age, gender, and control preferences, and challenge the assumption that instrumental qualities such as perceived usefulness and effectiveness uniformly favour task-oriented interactions. Although grounded in recommendation dialogues, the insights and adaptive heuristics developed here generalise to conversational AI systems that must dynamically align with users’ informational and affective needs.

The remainder of this article is structured as follows. Section 2 reviews related work. Section 3 outlines the study design. Section 4 reports the results. Section 5 discusses implications and limitations. Section 6 concludes.

\section{Related Work}
This section synthesises prior research that informs our study. We organise the review into four strands: (1) interaction styles, (2) CRS qualities, (3) UX dimensions, and (4) user profiling. Each strand contributes to understanding how users perceive and prefer different styles of conversational interaction. The first subsection addresses high-level dialogue strategies.

\subsection{Interaction Styles}
\label{subsec:interaction-styles}
While early work prioritised algorithmic improvements, recent studies have begun examining how interaction strategies affect user experience \cite{sun2018conversational, jannach2021survey}. Among these, the contrast between \emph{task-oriented} and \emph{exploratory} styles has gained particular attention. In this context, \emph{task-oriented} interactions aim to fulfil user goals efficiently, using brief exchanges that minimise user effort and reduce cognitive load. These systems are particularly suited to users with well-defined intents, and are often evaluated based on completion rates, decision time, and transactional efficiency \cite{zhang2020recent}.

In contrast, \emph{exploratory} interactions promote discovery by offering elaborative prompts, contextual suggestions, or reflective questions. They aim not only to deliver recommendations but also to surface latent needs, help users articulate vague or evolving goals, and refine preference structures through guided conversation. These strategies are rooted in principles from exploratory search \cite{marchionini2006exploratory, white2009exploratory}, preference elicitation \cite{pommeranz2012designing, priyogi2019preference}, and sensemaking theory \cite{pirolli2011introduction, russell1993cost}. For example, a CRS assisting with laptop shopping might, in a \emph{task-oriented} mode, list top-rated models after a single query specifying price or brand. In contrast, the same system operating in an \emph{exploratory} mode might begin by asking whether the user intends to use the device for gaming, remote work, or academic purposes, and follow up with questions about screen preferences, past frustrations, or anticipated mobility needs. Similarly, in a travel domain, a task-oriented CRS might immediately show flight deals based on a destination query, whereas an exploratory system could suggest alternative destinations aligned with the user’s mood, budget, or prior trips, then offer itinerary variations with local experiences. In education or upskilling contexts, an exploratory CRS might help users clarify goals by probing into their current career situation, prior coursework, or preferred learning style before suggesting a curated learning path. Through such layered exchanges, \emph{exploratory} interactions scaffold user decision-making in ways that are reflective, adaptive, and context-sensitive.

Although both styles have merits, they afford different experiential qualities. Task-oriented interactions typically yield speed, clarity, and transactional utility. Exploratory dialogues, while slower, support agency, reflection, and satisfaction in ill-structured decision tasks. 
Recent work has also introduced adaptive flows that adjust initiative based on user traits \cite{rieser2011reinforcement, ma2023initiative, du2024user}, but systematic modelling of how interaction preferences relate to user traits remains limited. Vakulenko et al. \cite{vakulenko2017conversational} proposed interactive storytelling to support exploratory search, while Lei et al. \cite{lei2020interactive} introduced the Estimation–Action–Reflection framework to facilitate adaptive flows. Gao et al. \cite{gao2021advances} further emphasised the importance of balancing exploration and exploitation in CRS design.
Despite these contributions, key questions remain about how perceived system qualities influence users’ interaction preferences. The next section examines these qualities in detail.

\subsection{CRS Qualities}
Beyond recommendation accuracy, users evaluate how well a system explains its suggestions, adapts to feedback, maintains conversational flow, and introduces novel or unexpected options. These perceptions influence both expectations and satisfaction \cite{pu2012evaluating, knijnenburg2012explaining}. The CRS-Que framework \cite{jin2024crs} builds on the earlier ResQue model \cite{pu2011user} by adapting its constructs specifically to conversational contexts, where dialogue structure and agent behaviour are critical. Key qualities identified include accuracy, novelty, explanation clarity, adaptability, and conversational quality \cite{jin2024crs}.
Unlike momentary user experience reactions, these qualities represent user beliefs about what the system should deliver, rather than transient feelings during use. Prior research shows that users’ preference of these qualities varies by domain and individual traits. For example, novelty and adaptability tend to be more valued in \emph{exploratory} or hedonic contexts, while accuracy and efficiency are paramount in \emph{task-oriented} contexts \cite{jannach2023evaluating, knijnenburg2012explaining}. Such distinctions suggest that CRS qualities are strong predictors of users’ dialogue strategy choices, especially when users must choose between \emph{task-oriented} and \emph{exploratory} dialogue. Moreover, although Berkovsky et al. \cite{berkovsky2008mediation} studied user model mediation and personalisation techniques primarily in broader recommender system domains rather than CRSs specifically, their work provides foundational methods for integrating heterogeneous user data sources. Such integration supports dynamic adaptation of system qualities and behaviours to reflect individual differences. This capacity for adaptation is crucial for tailoring interactions that align with user preferences and contexts in CRSs. However, system qualities alone do not fully explain how users appraise their experiences. UX research shows that subjective evaluations, especially affective and cognitive appraisals, play a crucial role.

\subsection{UX Dimensions}
In this study, UX refers specifically to post-interaction appraisals that reflect users’ cognitive and emotional evaluations of the experience. These appraisals are distinct from behavioural indicators such as time spent or action frequency. While behavioural data offer observable metrics of engagement, perceptual evaluations reveal underlying judgments shaped by users’ expectations, system behaviour, and contextual goals \cite{mccarthy2004technology, law2009understanding, o2010development}.

UX perceptions influence outcomes such as trust formation, decision confidence, and continued system acceptance \cite{bickmore2005social}. Affective dimensions such as enjoyment and surprise are particularly relevant in open-ended or elaborative conversations, where engagement and perceived intelligence of the system are crucial \cite{pramod2022conversational, deldjoo2404review}. 
In contrast, dimensions such as usefulness and perceived effectiveness are more salient in task-oriented interactions, reflecting the system’s ability to provide practical support and relevant outcomes \cite{konstan2012recommender}. Recent findings by Yun et al. \cite{yun2025user} show that affective tone and perceived helpfulness significantly influence interaction preferences in LLM-powered CRS dialogues, especially when users seek elaborative or emotionally intelligent responses. Similarly, Gajos et al. \cite{gajos2008automatically} demonstrate that adaptive interface behaviours tailored to user ability and context can enhance perceived control and user satisfaction, suggesting that adaptivity plays a critical role in shaping post-interaction UX.

Although UX is typically measured after interaction, it is not merely a passive outcome. Rather, it constitutes a structured, reflective judgment that mediates preference formation and behavioural intention \cite{gregor1999explanations, forlizzi2004understanding}. These post-task evaluations serve as interpretive summaries that integrate momentary affect with broader task-related reasoning. Kocaballi et al. \cite{kocaballi2019personalization} provide a comprehensive taxonomy of UX factors in conversational systems, including emotion, hedonic quality, motivation, and frustration. Their work highlights the multidimensionality of UX and the importance of capturing both experiential and evaluative components. Consistent with this view, the present study treats UX perceptions as explanatory variables in preference modelling, testing whether users’ retrospective evaluations help forecast their interaction preferences beyond what can be inferred from system quality priorities alone. These perceptual outcomes are also shaped by user-level traits. Thus, the next section turns to individual differences and profiling approaches in CRS design.

\subsection{User Profiling}
\label{par:user-profiling}
User preferences in CRSs are shaped by a range of individual differences, including demographic traits, cognitive styles, behavioural tendencies, and control expectations. A key concept in this context is value orientations, defined as stable patterns in users’ priorities and expectations regarding system attributes and interaction styles. These orientations shape how users evaluate and respond to different CRS behaviours. Research indicates that grouping users by their attitudes, behaviours, and value orientations helps tailor system strategies more effectively to meet individual needs \cite{tsai2021effects, purificato2024user, lakiotaki2011multicriteria}. For example, some users prioritise novelty and detailed explanations, shaping their expectations about dialogue style and system initiative \cite{siro2023understanding}. Beyond segmentation, individual characteristics also moderate how users respond to system features. Users with high digital agency or strong trust in technology often prefer adaptive, \emph{exploratory} interactions, whereas those who favour clear, directive guidance tend to prefer \emph{task-oriented} interaction. Aligning interaction patterns with these preferences is critical for user satisfaction and system acceptance \cite{tintarev2012evaluating}. Additionally, demographic factors such as age, language proficiency, and prior experience influence perceptions of agent tone, fluency, and usefulness \cite{folstad2020users, hornbaek2017technology}. Notably, Berkovsky et al. \cite{berkovsky2008mediation, berkovsky2012influencing, Berkovsky2016} demonstrate the benefits of integrating diverse user data sources and employing persuasive personalisation techniques to enhance recommendation relevance and user satisfaction. Their work supports the development of adaptive user profiles that enable personalised and context-aware CRS interactions. 

Collectively, these four strands of research underscore the complexity of CRS evaluation and personalisation. While prior work has explored dialogue strategies, system attributes, UX outcomes, and user traits, few studies have integrated these components to predict how individuals choose between dialogue styles. Moreover, existing frameworks tend to focus on retrospective evaluation rather than preference forecasting. By combining CRS qualities, retrospective UX ratings, and trait-level profiling in a predictive model, our study advances a more adaptive and explanatory perspective on dialogue preference in LLM-powered CRS environments.

\section{Methods}
\label{sec:methods}

We employed a within-subjects experimental design in which participants sequentially experienced two scripted CRS dialogues—one \emph{task-oriented} and one \emph{exploratory} as described in Section~\ref{subsec:interaction-styles}. The dialogue excerpts are given in Appendix~\ref{sec:scripted_dialogues}. The dialogues were situated in an online apparel-shopping context and were carefully matched for informational content and length to isolate the stylistic contrast. Participants rated their experiences, prioritised eight CRS qualities, and provided individual-difference data, allowing us to model preferences based on system quality beliefs, UX appraisals, and demographic and usage traits.  

\subsection{Research Design}
\label{sec:research-design}

The defining characteristics of the \emph{task-oriented} and \emph{exploratory} interaction styles are summarised in Table~\ref{tab:interaction_modes}. These styles were operationalised based on exploratory search theory \cite{marchionini2006exploratory}, conversational dialogue design principles \cite{rieser2011reinforcement}, and the CRS-Que framework \cite{jin2024crs}, which together provide theoretical grounding for contrasting interaction modes in CRS dialogues. The use of scripted dialogues was a deliberate methodological choice to maintain experimental control and stimulus equivalence, enabling systematic examination of dialogue style effects on user preference. Scripted or synthetic interactions are widely employed in conversational AI and human–computer interaction research to facilitate within-subject comparisons while controlling for content and length, thereby reducing variability inherent in live interactions \cite{bickmore2005social, busso2008iemocap, rieser2011reinforcement}. Although scripted dialogues do not capture dynamic turn-taking or adaptive responses, their controlled nature enhances internal validity and supports precise measurement of user evaluations \cite{chaves2021should, mctear2016conversational}.
After exposure to both dialogue conditions, participants indicated their preferred interaction style and completed survey measures assessing prioritised system qualities, retrospective user experience evaluations, and individual characteristics. To validate these scripts, we conducted pilot testing with a small sample and solicited expert feedback. Reviewers confirmed that the \emph{task-oriented} dialogue conveyed a goal-focused concise tone, while the \emph{exploratory} version was perceived as elaborative and suggestion-driven.

\begin{table}[h]
  \caption{Key characteristics of the \emph{task-oriented} and \emph{exploratory} interaction styles used in the scripted CRS dialogues.}
  \label{tab:interaction_modes}
  \centering
  \small
  \begin{tabular}{p{2.3cm} p{5cm} p{5.5cm}}
    \toprule
    \textbf{Aspect} & \textbf{Task-oriented} & \textbf{Exploratory} \\
    \midrule
    Dialogue style         & Concise, direct, goal-oriented            & Elaborative, open-ended, suggestion-rich   \\
    System prompts         & Minimal elaboration, rapid delivery       & Reflective, contextual, inviting elaboration \\
    User engagement        & Decision-focused, minimal probing         & Exploration-focused, encourages deliberation \\
    Tone                   & Functional, task-driven                   & Conversational, empathic, exploratory       \\
    System behaviours & Highlights deals, delivers options quickly & Offers context, rationale and alternatives  \\
    \bottomrule
  \end{tabular}
\end{table}

\subsection{Participant Recruitment and Data Collection}
Participants (\(N = 191\)) were recruited from three online platforms—Prolific (\(n = 149\)), Survey Swap (\(n = 33\)), and Meta (\(n = 9\))—to ensure a demographically diverse sample. Eligibility criteria required participants to be at least 18 years old and proficient in English. For Prolific recruits, this was verified through platform-level screening (minimum 95\% approval rate and at least 100 prior completions); for other sources, proficiency was self-reported. No restrictions were placed on gender, education level, or geographic location.
Participants completed a structured Qualtrics survey lasting approximately nine minutes and received \$2.40 as compensation. To ensure data quality, we applied a multi-criteria screening protocol that excluded responses with unusually fast completion times, straightlining patterns, or excessive missing or inconsistent data \cite{meade2012identifying, laranjeiro2015survey}.
Following screening, \(N = 168\) valid responses remained. Of these, \(N = 139\) participants expressed a definitive interaction preference and were retained for inferential analysis, a sample size that satisfies recommended thresholds for statistical power in logistic regression and unsupervised clustering \cite{hosmer2013applied, fraley1998many}. No personally identifiable information was collected, ensuring full compliance with institutional ethics and data protection regulations.

\vspace{-4pt}
\subsection{Survey Instruments and Measures}
The survey consisted of six modules: participant consent, demographics, dialogue stimuli exposure, UX appraisals, CRS quality prioritisation, and usage-related variables. Demographic data included age group, gender, education level, occupation, and self-reported English language proficiency. As described in Section~\ref{sec:research-design}, participants experienced two matched scripted CRS dialogues differing only in interaction style (\emph{task-oriented} vs. \emph{exploratory}), allowing us to isolate stylistic effects.
Interaction preference was assessed using a categorical forced-choice item, where participants indicated which of the two dialogues they found more desirable or preferable. The response options included preference for Dialogue 1 (task-oriented), Dialogue 2 (exploratory), both equally, none, or unsure. For analytical clarity, only participants indicating a clear preference for either Dialogue 1 or 2 were retained. These responses were recoded into a binary variable representing preference for the \emph{task-oriented} (coded 0) or \emph{exploratory} (coded 1) dialogue style. Neutral or ambivalent responses (both, none, unsure) were excluded from inferential analyses.

User experience was evaluated across four dimensions: enjoyment, surprise, usefulness, and perceived effectiveness. These dimensions were adapted from established human–computer interaction frameworks \cite{hassenzahl2010experience, knijnenburg2012explaining} and were rated separately for each dialogue. Although measured on Likert scales, UX scores were treated as continuous variables and standardised prior to analysis, consistent with accepted psychometric practice \cite{norman2010likert, carifio2008resolving}.

Participants also rated the importance of eight CRS qualities, derived from the CRS-Que framework \cite{jin2024crs}, including accuracy, novelty, explanation clarity, adaptability, conversation quality, attentiveness, understanding, and response quality. These ratings served as independent predictors in regression analyses and as features for clustering models. All CRS quality ratings were standardised using z-scores.
Participants reported their CRS usage frequency on an ordinal scale ranging from ‘Never’ to ‘Always’. Control preferences regarding system-initiated versus user-initiated interactions were measured using five-point Likert-type items. For regression and moderation analyses, these variables were treated as continuous predictors following standard practice \cite{norman2010likert}.

\subsection{Data Analysis}

All analyses were conducted in IBM SPSS and Python using pandas, statsmodels, and scikit-learn. Predictor variables measured on five-point Likert scales were standardised to z-scores prior to analysis to place them on a common metric and facilitate interpretation. 

\paragraph{RQ1: Binary Logistic Regression}
To model the probability that participant \(i\) preferred the exploratory interaction (\(Y_i = 1\)) over the task-oriented interaction (\(Y_i = 0\)), we applied binary logistic regression. This method is suitable for dichotomous dependent variables and enables interpretation of predictor effects in terms of log-odds, which is commonly used in behavioural research \cite{hosmer2013applied}.

\begin{equation}
\mathrm{logit}\bigl(P(Y_i = 1)\bigr)
= \beta_{0} + \sum_{j=1}^{k}\beta_{j}\,x_{ij},
\label{eq:logistic}
\end{equation}

where \(\mathrm{logit}(P)\) denotes the natural logarithm of the odds of preferring the exploratory interaction, \(\beta_{0}\) is the model intercept, \(\beta_{j}\) are regression coefficients, and \(x_{ij}\) are the standardised predictor variables (e.g., CRS qualities and UX dimensions).

\paragraph{RQ2: Unsupervised Clustering}
To identify latent user profiles, we applied k-means clustering to participants’ standardised CRS quality ratings. This technique partitions users into \(K\) clusters by minimising within-cluster variance:

\begin{equation}
\min_{\{\mu_k\}}
\sum_{i=1}^{N}\sum_{k=1}^{K}
r_{ik}\,\lVert x_{i} - \mu_{k}\rVert^{2},
\quad
r_{ik} =
\begin{cases}
1 & \text{if participant } i \text{ is assigned to cluster } k,\\
0 & \text{otherwise},
\end{cases}
\label{eq:kmeans}
\end{equation}

where \(\mu_k\) are cluster centroids. We then validated and refined this solution via hierarchical agglomerative clustering using Ward’s method, which iteratively merges clusters to minimize increases in total within-cluster variance. Candidate values of \(K\) from 3 to 9 were evaluated using the elbow and silhouette criteria and assessed for stability via perturbation-bootstrap adjusted Rand Index.

\paragraph{RQ3: Moderated Logistic Regression}
To examine whether user traits (e.g., age, control preference) moderated the influence of CRS predictors on interaction preference, we extended the logistic model to include interaction terms. This model tests whether the effect of each CRS quality \(x_{ij}\) on interaction preference is moderated by a corresponding user trait \(z_{im}\), through the inclusion of interaction terms \((x_{ij} \times z_{im})\).

\begin{equation}
\mathrm{logit}\bigl(P(Y_i = 1)\bigr) = \beta_{0} + \sum_{j=1}^k \beta_{j} x_{ij} + \sum_{m=1}^M \gamma_{m} z_{im} + \sum_{j=1}^k \sum_{m=1}^M \delta_{jm}\,(x_{ij} \times z_{im}),
\label{eq:moderated}
\end{equation}

where \(x_{ij}\) are CRS predictors, \(z_{im}\) are user trait moderators, \(\beta_j\) and \(\gamma_m\) represent main effects, and \(\delta_{jm}\) are coefficients for the interaction terms. Interaction effects were visualised using marginal-effects plots and stratified models where appropriate.

\paragraph{Hypothesis Testing Criteria}
Hypotheses were evaluated using a two‐tailed significance threshold of \(p \leq 0.05\), consistent with standard practice in behavioural research \cite{cohen1992power, field2024discovering}. A result yielding \(p \leq 0.05\) was taken as sufficient evidence to reject the null hypothesis \(H_{0}\) in favour of the corresponding alternative hypothesis \(H_{1}\). Conversely, a result with \(p > 0.05\) was interpreted as indicating insufficient evidence to reject \(H_{0}\), without implying proof of its truth. This approach ensures that Type I error is controlled at \(\alpha = 0.05\) while acknowledging that failure to reject \(H_{0}\) reflects only a lack of decisive evidence against it given the sample data \cite{wasserstein2016asa}.  

\subsection{Research Hypotheses}

\paragraph{RQ1: Predictive Role of CRS Qualities and UX Dimensions}
The following hypotheses were preregistered to assess how system‐level beliefs and retrospective experience appraisals predict interaction preference. Hypothesis H1a states that users’ preference for specific CRS qualities (for example, novelty and adaptability) predicts their likelihood of preferring the \emph{exploratory} interaction. Hypothesis H1b asserts that users’ preference for qualities related to accuracy and response quality predicts their likelihood of preferring the \emph{task-oriented} interaction. Hypothesis H1c proposes that higher retrospective ratings of enjoyment and surprise are positively associated with preference for the \emph{exploratory} interaction. Finally, Hypothesis H1d posits that higher retrospective ratings of usefulness and perceived effectiveness are positively associated with preference for the \emph{task-oriented} interaction.

\paragraph{RQ2: Latent User Profiles Based on Value Orientations}
To investigate whether distinct user segments emerge from differential value orientations, Hypothesis H2a proposes that participants can be clustered into latent profiles according to their prioritised CRS qualities. Hypothesis H2b further predicts that these latent profiles will exhibit systematic differences in interaction preference.

\paragraph{RQ3: Moderation by Usage Patterns and Demographic Traits}
The final set of hypotheses examines whether individual characteristics moderate the influence of CRS qualities on interaction preference. Hypothesis H3a predicts that CRS usage frequency will moderate the relationship between CRS qualities and preferred interaction style. Hypothesis H3b asserts that control preference (system‐initiated versus user‐initiated) will serve as a moderator of the same relationship. Hypothesis H3c proposes that demographic factors, specifically age and gender, will moderate users’ interaction preferences in response to varying CRS qualities.

\subsection{Ethics}
The study protocol was reviewed and approved by the Human Research Ethics Committee of the University of Technology Sydney. All participants provided informed consent prior to participation. The study adhered to institutional guidelines for ethical research involving human subjects and complied with relevant data protection regulations, including the General Data Protection Regulation (GDPR). Participants’ privacy and confidentiality were rigorously maintained. No personally identifiable information, such as names, geographic locations, or sensitive demographic details, was collected. Data were anonymised and stored securely in accordance with the University’s data management policies.

\section{Results}
We organised this section sequentially to align with each research question and its associated hypotheses, beginning with predictive models of interaction preference (RQ1), followed by cluster-based user profiling (RQ2), and concluding with moderation analyses exploring the influence of individual characteristics (RQ3). A consolidated summary presenting the key results is provided in Table~\ref{tab:key_findings}.

\subsection{Predictors of Interaction Preference (RQ1)}
Two binary logistic regression models were fitted: one using eight CRS quality attributes derived from the CRS-Que framework \cite{jin2024crs}, and another using four UX dimensions—enjoyment, surprise, usefulness, and perceived effectiveness, as expressed in Equation~\ref{eq:logistic}. Both models were statistically significant, with McFadden’s pseudo \(R^2\) values of 0.12 and 0.32, respectively, indicating modest to moderate explanatory power.
As shown in Table~\ref{tab:logreg_crs_qualities}, higher preference for \emph{novelty} (Odds Ratio [OR] = 1.66, 95\% Confidence Interval [CI] [1.04, 2.63], \(p = 0.033\)) and \emph{conversation quality} (OR = 1.77, 95\% CI [1.04, 2.99], \(p = 0.035\)) significantly increased the odds of preferring the \emph{exploratory} interaction. Preference for \emph{accuracy} (OR = 1.11, \(p = 0.674\)) and \emph{response quality} (OR = 0.79, \(p = 0.383\)) did not significantly predict preference for the \emph{task-oriented} interaction. Explainability showed a trend toward significance (\(p = 0.062\)), while attentiveness, understanding, and adaptability were not significant predictors.

\begin{table}[htbp]
  \caption{Logistic regression model predicting preference for \emph{exploratory} interaction based on preferred CRS qualities.  
  \textit{Note.} OR > 1 indicates greater odds of preferring the \emph{exploratory} interaction. Values in bold denote statistical significance at \(p \leq 0.05\). McFadden’s pseudo \(R^2 = 0.12\); \(N = 139\).}
  \label{tab:logreg_crs_qualities}
  \centering
  \small
  \begin{tabular}{lccc}
    \toprule
    \textbf{Predictor} & \textbf{Odds ratio (OR)} & \textbf{95\% CI [lower, upper]} & \textbf{$p$-value} \\
    \midrule
    \multicolumn{4}{l}{\textit{H1a: Predicted positive association with \emph{exploratory} interaction}} \\
    Novelty               & 1.66 & [1.04, 2.63] & \textbf{.033} \\
    Adaptability          & 0.79 & [0.48, 1.31] & .366 \\
    Attentiveness         & 1.00 & [0.59, 1.67] & .988 \\
    \midrule
    \multicolumn{4}{l}{\textit{H1b: Predicted positive association with \emph{task-oriented} interaction}} \\
    Accuracy              & 1.11 & [0.69, 1.77] & .674 \\
    Response quality      & 0.79 & [0.47, 1.34] & .383 \\
    \midrule
    \multicolumn{4}{l}{\textit{Other CRS qualities (included for completeness)}} \\
    Explainability        & 1.57 & [0.98, 2.52] & .062 \\
    Conversation quality  & 1.77 & [1.04, 2.99] & \textbf{.035} \\
    Understanding         & 0.70 & [0.42, 1.16] & .163 \\
    \bottomrule
  \end{tabular}
\end{table}

Table~\ref{tab:logreg_ux_dimensions} presents results from the UX dimension model. Higher ratings of \emph{enjoyment} (OR = 2.19, 95\% CI [1.37, 3.51], \(p = 0.001\)), \emph{usefulness} (OR = 2.23, 95\% CI [1.32, 3.77], \(p = 0.003\)), and \emph{perceived effectiveness} (OR = 2.17, 95\% CI [1.28, 3.69], \(p = 0.004\)) were significantly associated with increased odds of preferring the \emph{exploratory} interaction. The \emph{surprise} dimension was not a significant predictor (OR = 0.68, 95\% CI [0.42, 1.10], \(p = 0.119\)). These findings support Hypotheses H1a and H1c, indicating that users who prefer novelty, conversation quality, and positive affective UX components such as enjoyment are more likely to prefer the \emph{exploratory} interaction. Hypotheses H1b and H1d were not supported, as preference for accuracy and response quality did not predict preference for the \emph{task-oriented} interaction, and usefulness and perceived effectiveness unexpectedly related positively to \emph{exploratory} preference. In summary, participants’ preference for the \emph{exploratory} interaction was more strongly associated with \emph{novelty} and \emph{conversation quality} than with \emph{accuracy} or \emph{response quality}. These results indicate that, when controlling for interaction style, users place greater emphasis on hedonic and conversational qualities rather than solely on instrumental attributes.

\begin{table}[htbp]
  \caption{Logistic regression model predicting preference for \emph{exploratory} interaction based on UX dimensions.  
  \textit{Note.} OR > 1 indicates greater odds of preferring the \emph{exploratory} interaction. Values in bold denote statistical significance at \(p \leq 0.05\). McFadden’s pseudo \(R^2 = 0.32\); \(N = 139\).}
  \label{tab:logreg_ux_dimensions}
  \centering
  \small
  \begin{tabular}{lccc}
    \toprule
    \textbf{Predictor} & \textbf{Odds ratio (OR)} & \textbf{95\% CI [lower, upper]} & \textbf{$p$-value} \\
    \midrule
    \multicolumn{4}{l}{\textit{H1c: Affective UX dimensions}} \\
    Enjoyment               & 2.19 & [1.37, 3.51] & \textbf{.001} \\
    Surprise                & 0.68 & [0.42, 1.10] & .119 \\
    \midrule
    \multicolumn{4}{l}{\textit{H1d: Task-oriented UX dimensions}} \\
    Usefulness              & 2.23 & [1.32, 3.77] & \textbf{.003} \\
    Perceived effectiveness & 2.17 & [1.28, 3.69] & \textbf{.004} \\
    \bottomrule
  \end{tabular}
\end{table}

\subsection{Cluster Profiles and Interaction Preference (RQ2)}
\label{sec:results-cluster-profiles}

We applied both k-means and hierarchical (Ward’s method) clustering to participants’ z-standardised ratings of eight CRS-Que qualities. Although both methods yielded similar high-level segment structures, we selected the hierarchical solution for reporting because it offered superior stability and more balanced cluster sizes as shown in Appendix Figure~\ref{fig:appendix_rq2_clustering_diagnostics} and Table~\ref{tab:appendix_cluster_diagnostics} \cite{arbelaitz2013extensive}. Solutions for \(k=3\)–\(9\) were evaluated using combined elbow-method and silhouette-score diagnostics; the five-cluster solution (\(k=5\)) achieved the best trade-off among within-cluster cohesion, between-cluster separation, stability (mean ARI = 0.770, SD = 0.214), and minimum cluster size (\(n\ge18\)). All subsequent analyses focus on this five-cluster segmentation.
The association between cluster membership and interaction preference was robust for the hierarchical solution (\(\chi^2(4, N=139)=14.93\), \(p=0.0048\); Kruskal–Wallis \(H=14.82\), \(p=0.0051\)). The average silhouette score (0.146) further confirmed the validity of the segmentation.

\paragraph{Cluster Characterisation}
Table~\ref{tab:cluster_means_5} reports the mean importance ratings for each CRS quality by cluster. Each column represents one user segment; the header row indicates cluster size. Cluster 1 (\(n=34\)) consistently rated all qualities highly (means 4.41–4.91), while Cluster 2 (\(n=18\)) showed moderate to low engagement (means 2.89–3.33). Cluster 3 (\(n=24\)) prioritised adaptivity and response quality but rated novelty lower. Cluster 4 (\(n=26\)) valued accuracy and adaptivity with moderate novelty and conversation quality. Cluster 5 (\(n=37\)) emphasised novelty, adaptivity, and response quality, with balanced ratings elsewhere.

\begin{table}[htbp]
  \caption{Mean importance ratings for eight CRS qualities by cluster (hierarchical Ward’s method, \(k=5\), \(N=139\)). Scores on 1–5 scale (1 = low, 5 = high).}
  \label{tab:cluster_means_5}
  \centering
  \small
  \begin{tabular}{lccccc}
    \toprule
    \textbf{CRS Quality} & \textbf{C1 (\(n=34\))} & \textbf{C2 (\(n=18\))} & \textbf{C3 (\(n=24\))} & \textbf{C4 (\(n=26\))} & \textbf{C5 (\(n=37\))} \\
    \midrule
    Accuracy          & 4.85 & 3.22 & 4.17 & 4.81 & 3.95 \\
    Explanation       & 4.62 & 3.11 & 3.71 & 3.65 & 3.51 \\
    Novelty           & 4.41 & 3.11 & 2.71 & 3.54 & 4.03 \\
    Conv.\ Quality    & 4.79 & 3.22 & 2.96 & 3.81 & 3.92 \\
    Attentiveness     & 4.91 & 3.28 & 3.50 & 4.27 & 3.97 \\
    Understanding     & 4.74 & 3.17 & 3.96 & 4.77 & 4.03 \\
    Adaptability      & 4.68 & 2.89 & 4.29 & 4.19 & 4.32 \\
    Response Quality  & 4.91 & 3.33 & 3.96 & 4.46 & 4.22 \\
    \bottomrule
  \end{tabular}
\end{table}

\paragraph{Interaction Preference Patterns}
Table~\ref{tab:preference_by_cluster_5} shows how participants’ interaction preferences (0 = task-oriented, 1 = exploratory) distribute across clusters. Cluster 1 exhibited a strong exploratory preference (23 of 34 participants). Cluster 2 was balanced (9 of 18). Clusters 3, 4, and 5 each favoured the task-oriented mode: Cluster 3 (17 of 24), Cluster 4 (18 of 26), and Cluster 5 (26 of 37). For a graphical overview of cluster value profiles, see Appendix Figure~\ref{fig:rq2_cluster_profiles}.

\begin{table}
  \caption{Counts of task-oriented versus exploratory preferences by cluster (\(k=5\); \(N=139\)).}
  \label{tab:preference_by_cluster_5}
  \centering
  \small
  \begin{tabular}{lcc}
    \toprule
    \textbf{Cluster} & \textbf{Task-oriented (0)} & \textbf{Exploratory (1)} \\
    \midrule
    1 & 11 & 23 \\
    2 &  9 &  9 \\
    3 & 17 &  7 \\
    4 & 18 &  8 \\
    5 & 26 & 11 \\
    \bottomrule
  \end{tabular}
\end{table}

\subsection{Moderating Effects of User Characteristics (RQ3)}

\begin{figure}[htbp]
    \centering
    \begin{tabular}{cc}
        \includegraphics[width=0.45\linewidth]{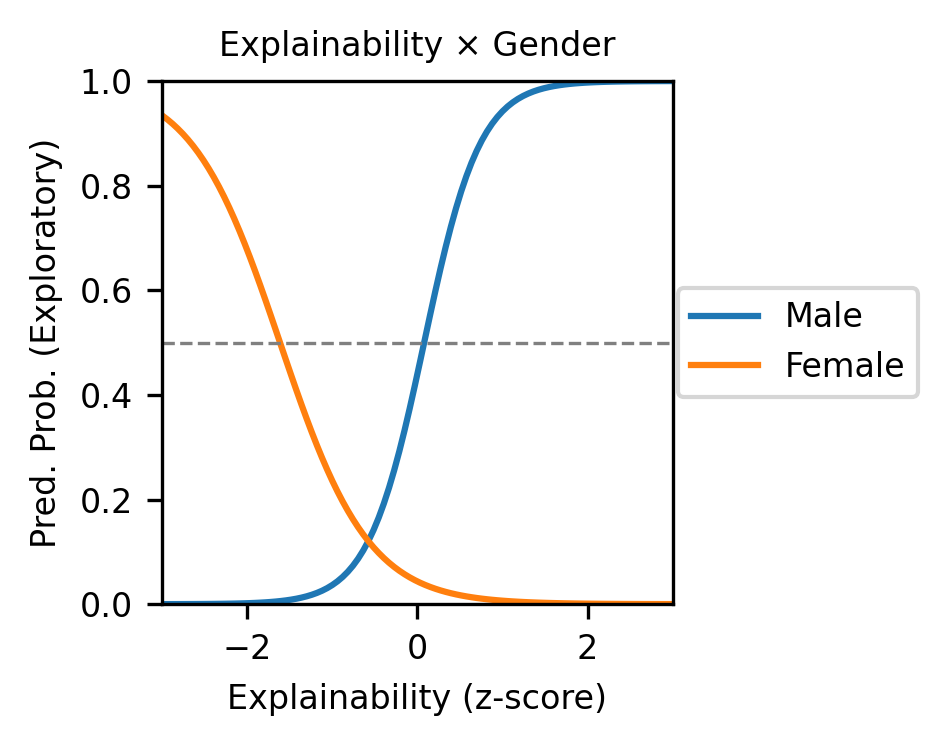} &
        \includegraphics[width=0.45\linewidth]{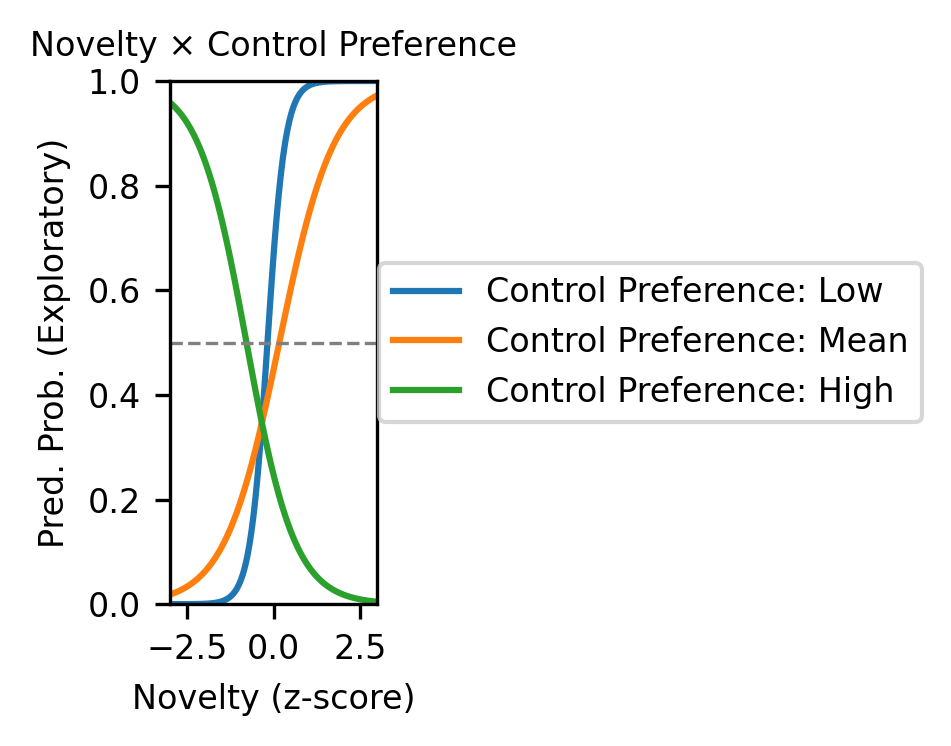} \\
        (a) Explainability $\times$ Gender & (b) Novelty $\times$ Control Preference \\
        \includegraphics[width=0.45\linewidth]{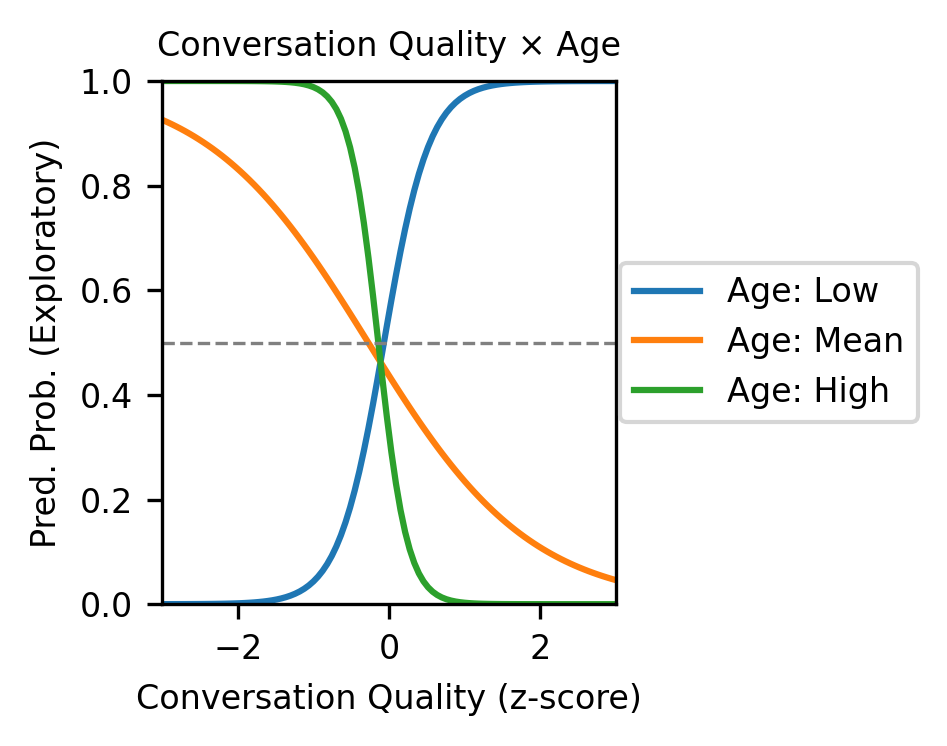} &
        \includegraphics[width=0.45\linewidth]{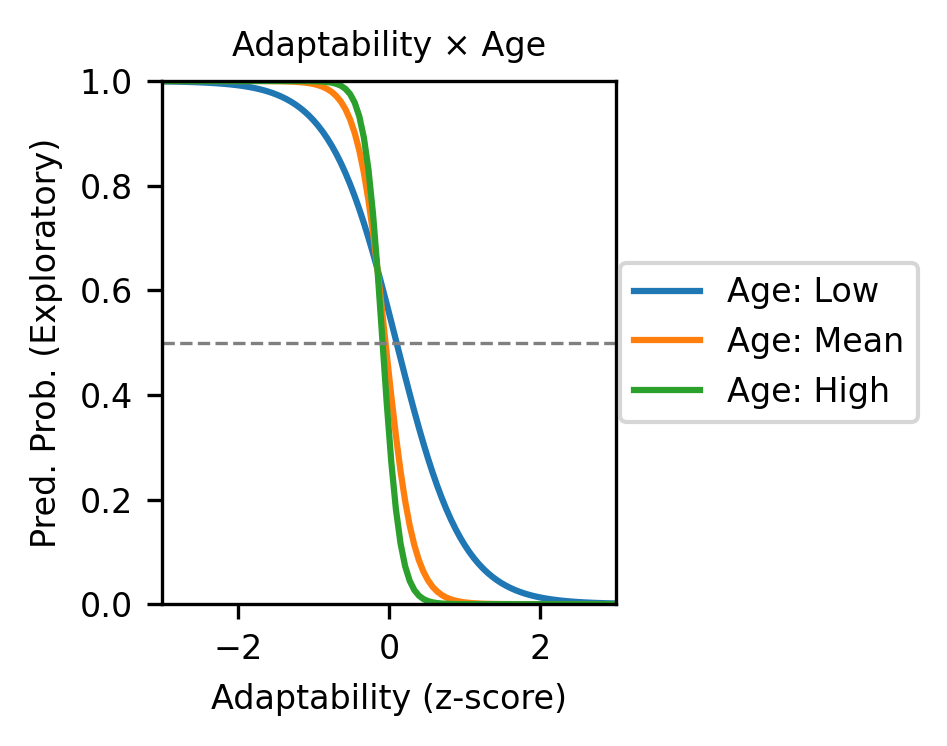} \\
        (c) Conversation Quality $\times$ Age & (d) Adaptability $\times$ Age \\
        \includegraphics[width=0.45\linewidth]{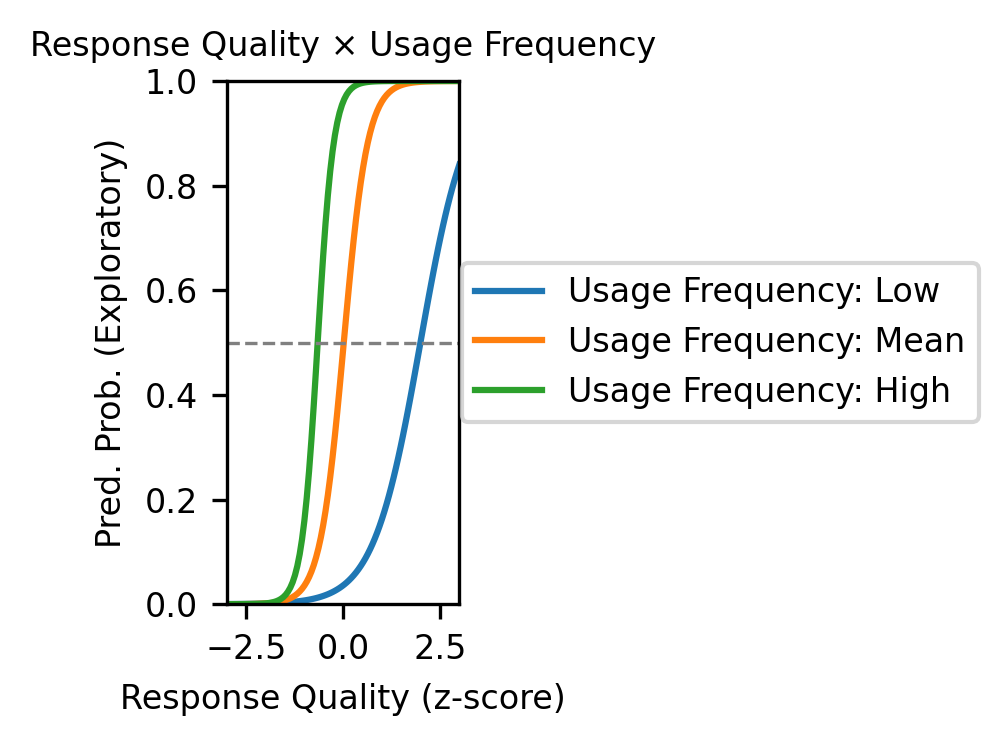} &
        \includegraphics[width=0.45\linewidth]{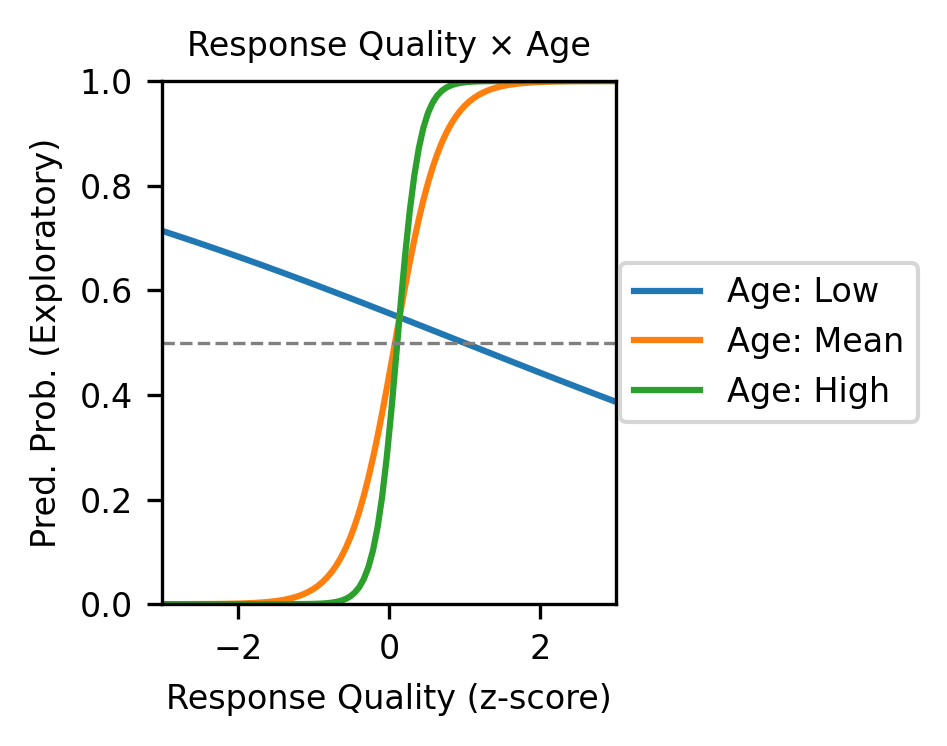} \\
        (e) Response Quality $\times$ Usage Frequency & (f) Response Quality $\times$ Age \\
    \end{tabular}
    \caption{
        Significant moderation effects for \emph{exploratory} interaction preference. Each panel shows the predicted probability of preferring the \emph{exploratory} interaction style as a function of CRS quality ($z$-score), stratified by levels of the moderator. Statistical details are provided in Table~\ref{tab:rq3_interactions}.
    }
    \label{fig:rq3_moderation_matrix}
\end{figure}

A series of moderated logistic regression analyses were conducted to examine whether the effects of perceived CRS qualities on interaction preference were systematically moderated by individual user characteristics, including usage frequency, control preference, age, and gender. The outcome variable was binary, reflecting preference for the \emph{exploratory} interaction style.
Eight CRS qualities were included as predictors, each standardised ($z$-scored). For each quality, interaction terms were specified with all four moderators. Significant interaction effects (\(p \leq 0.05\)) are summarised in Table~\ref{tab:rq3_interactions}. Six interaction terms were statistically significant at the \(p \leq 0.05\) level: Explainability $\times$ Gender, Novelty $\times$ Control Preference, Conversation Quality $\times$ Age, Adaptability $\times$ Age, Response Quality $\times$ Usage Frequency, and Response Quality $\times$ Age. For all significant interactions, model-predicted probabilities of preferring the \emph{exploratory} interaction style are visualised in Figure~\ref{fig:rq3_moderation_matrix}. In each panel, the probability of \emph{exploratory} interaction preference is plotted as a function of the corresponding CRS quality ($z$-score), stratified by levels of the moderating variable.
No other quality $\times$ moderator interactions reached statistical significance at \(p \leq 0.05\).

\begin{table}[h]
    \centering
    \caption{Significant interaction effects (\(p \leq 0.05\)) from moderated logistic regression predicting preference for the \emph{exploratory} interaction mode.}
    \label{tab:rq3_interactions}
    \small
    \begin{tabular}{p{3.5cm} p{3cm} r r r r}
        \toprule
        \textbf{CRS Quality} & \textbf{Moderator} & \textbf{Estimate} & \textbf{Std. Error} & \textbf{\(z\)} & \textbf{\(p\)} \\
        \midrule
        Explainability       & Gender           & $-4.94$ & $2.11$ & $-2.34$ & $\mathbf{.019}$ \\
        Novelty              & Control Pref.    & $2.79$  & $1.30$ & $2.14$  & $\mathbf{.033}$ \\
        Conversation Quality & Age              & $-4.00$ & $1.98$ & $-2.02$ & $\mathbf{.044}$ \\
        Adaptability         & Age              & $-2.95$ & $1.50$ & $-1.96$ & $\mathbf{.050}$ \\
        Response Quality     & Usage Frequency  & $1.53$  & $0.75$ & $2.04$  & $\mathbf{.041}$ \\
        Response Quality     & Age              & $3.28$  & $1.37$ & $2.40$  & $\mathbf{.016}$ \\
        \bottomrule
    \end{tabular}
\end{table}

\subsection{Summary of Hypotheses Tests}
\label{sec:summary_findings}
We summarise the outcomes of all nine preregistered hypotheses associated with RQ1 through RQ3. Table~\ref{tab:key_findings} presents each hypothesis alongside its empirical outcome and statistical evidence, offering a concise overview of which predictions were supported, partially supported, or not confirmed. Notably, hypotheses concerning novelty (H1a), as well as user clustering (H2a–H2b), received strong empirical support. In contrast, predictions grounded solely in accuracy (H1b) were not supported. Moderation effects emerged for usage frequency, control preference, age, and gender (H3a–H3c), highlighting the contextual conditions under which system quality beliefs influence interaction preferences.

\begin{table}[h]
\caption{Summary of hypothesis outcomes and supporting statistical evidence for RQ1–RQ3. \textit{Note.} OR > 1 indicates greater odds of preferring the \emph{exploratory} interaction; statistical significance at \(p \leq 0.05\).}
\label{tab:key_findings}
\centering
\small
\begin{tabular}{p{5.2cm} p{2.6cm} p{5.5cm}}
\toprule
\textbf{Hypothesis Statement} & \textbf{Empirical Outcome} & \textbf{Statistical Evidence} \\
\midrule
\textbf{H1a.} Preference for novelty or adaptability predicts exploratory interaction.
  & Partially supported
  & Novelty: OR = 1.66, \(p = .033\); Adaptability: OR = .79, \(p = .366\) \\

\textbf{H1b.} Preference for accuracy and response quality predicts task-oriented interaction.
  & Not supported
  & Accuracy: \(p = .674\); Response quality: \(p = .383\) \\

\textbf{H1c.} Enjoyment and surprise predict exploratory preference.
  & Partially supported
  & Enjoyment: OR = 2.19, \(p = .001\); Surprise: \(p = .119\) \\

\textbf{H1d.} Usefulness and perceived effectiveness predict task-oriented preference.
  & Refuted
  & Usefulness: OR = 2.23, \(p = .003\); Perceived effectiveness: OR = 2.17, \(p = .004\) \\

\textbf{H2a.} Users can be clustered by preferred CRS qualities.
  & Supported
  & $k=5$ solution: silhouette = 0.146; ARI = 0.770 (SD = 0.214); elbow inflection at $k=5$ \\

\textbf{H2b.} Clusters differ in interaction preference.
  & Supported
  & $\chi^2(4, N=139)=10.47$, $p=0.033$; Kruskal–Wallis $H=10.40$, $p=0.034$ \\

\textbf{H3a.} CRS usage frequency moderates CRS quality\(\times\)preference.
  & Supported
  & Response Quality \(\times\) Usage Frequency: \(p = .041\) \\

\textbf{H3b.} Control preference moderates CRS quality\(\times\)preference.
  & Supported
  & Novelty \(\times\) Control Preference: \(p = .033\) \\

\textbf{H3c.} Age and gender moderate CRS quality\(\times\)preference.
  & Supported
  & Explainability \(\times\) Gender: \(p = .019\); Conversation Quality \(\times\) Age: \(p = .044\); Adaptability \(\times\) Age: \(p = .050\); Response Quality \(\times\) Age: \(p = .016\) \\
\bottomrule
\end{tabular}
\end{table}

\section{Discussion}

The findings advance the understanding of how user-valued system qualities, affective experiences, and individual traits dynamically converge to shape dialogue choice in CRS. We organised these findings around our three research questions to draw out both theoretical contributions and practical recommendations.

\subsection{Key Insights}

\textbf{First}, echoing work on hedonic system attributes in e-commerce and multimedia recommendation \cite{zhou2020improving, zhang2024navigating, castells2021novelty, kossack2023emotion}, novelty and conversational quality emerged as robust predictors of preference for the \emph{explorative} dialogue. Functional attributes such as accuracy and response quality, by contrast, showed limited predictive value when participants evaluated the dialogue styles side by side. This reinforces findings that affective and experiential dimensions often outweigh pure performance metrics in AI interactions \cite{kim2025enhancing, lalmas2022measuring, bilquise2022emotionally, mitopenai2025affective, zhou2020design}.
\textbf{Second}, beyond these experiential drivers, perceived usefulness also forecast selection of the \emph{explorative} style. This suggests that participants equated usefulness with informational depth and clarity, features more characteristic of open-ended exchanges than terse, \emph{task-oriented} ones, and underscores the role of dialogue richness in user sensemaking \cite{pirolli2005sensemaking, savolainen1993sense, dervin1998sense}.
\textbf{Third}, our clustering uncovered five latent profiles based on quality priorities, each showing distinct style preferences. This validates value-based segmentation as a cold-start personalisation strategy when behavioural histories are sparse
\cite{knijnenburg2012explaining, tintarev2015explaining, hattori2014recommender, jannach2023survey}.
\textbf{Fourth}, moderation tests revealed that age, gender, and control expectations shape how quality beliefs translate into style choices. For instance, conversational clarity carried more weight for older users, while explainability effects differed by gender. These nuances argue for adaptive systems that tailor both content and autonomy levels to user traits \cite{schemmer2023towards, ma2024investigating, laban2020effect, wu2025negotiating, del2024ai}.
These insights pave the way for the theoretical elaborations and concrete design patterns that follow.

\subsection{Theoretical and Design Implications}

The findings contribute to theory by reframing interaction preference as a predictive construct influenced by system quality beliefs, retrospective UX appraisals, and individual differences. Building on the original CRS-Que framework \cite{jin2024crs}, which treated system qualities solely as evaluation metrics, we demonstrate their forward-looking power in forecasting dialogue choice. In particular, novelty, conversational quality, and perceived usefulness, dimensions identified as user-valued, emerged as significant predictors of preference for \emph{exploratory} interaction. This reinforces HCI theories that emphasise the formative role of experiential qualities in shaping expectations and behaviours through cognitive sensemaking and affective engagement \cite{pirolli2005sensemaking, stige2024artificial, rana2024user, xu2024ai}.

Unexpectedly, hypothesis H1d, which predicted that higher ratings of usefulness and perceived effectiveness would favour the \emph{task-oriented} mode, was refuted. Instead, participants construed these instrumental dimensions in terms of informational completeness and clarity, aligning them more closely with \emph{exploratory} dialogues than with terse exchanges. Consequently, \emph{exploratory} interactions may be perceived as more “effective” because they provide essential context and explanations for comprehension and decision-making. This insight highlights the inherently dual cognitive and affective character of usefulness and perceived effectiveness, and suggests that genuine efficiency requires design innovations such as adaptive summarisation or information-density controls. These mechanisms preserve perceived utility while streamlining dialogue.
Taken together, these theoretical advances point to a refined model of CRS adaptation, in which both experiential and instrumental dimensions jointly inform dynamic dialogue strategies.

Complementing these contributions, our identification of five latent user profiles based on quality-preference clustering offers empirical support for value-based segmentation in cold-start scenarios \cite{knijnenburg2012explaining, knijnenburg2015evaluating, de2023systematic}. Each profile exhibited distinctive interaction preferences, validating segmentation heuristics that anticipate user expectations from stated priorities alone. Moderation analyses further extend autonomy-centred design paradigms by revealing how trait-level differences, such as age, gender, and control expectations, shape the weight users assign to specific system qualities when choosing between interaction styles \cite{singh2024recommender, del2024ai, krook2023recommender, bonicalzi2023artificial}.
From a practical standpoint, these insights prescribe a new class of adaptive dialogue management systems. Specifically, designers should implement threshold-based switching rules that monitor real-time indicators (e.g.\ novelty sensitivity, enjoyment, control preference) to determine when to transition between task-oriented and exploratory strategies. Such heuristics reconcile diverse user needs, enabling systems to modulate tone, elaboration, and initiative according to inferred user values and traits.

Collectively, these theoretical and design implications chart a path toward next-generation CRS architectures that transcend traditional performance metrics such as accuracy, precision, and recall. They integrate predictive preference modelling, robust value-profile segmentation, moderation-guided personalisation, and dynamic adaptation heuristics. In doing so, they lay the groundwork for conversational experiences that are truly user-centred, context-sensitive, and capable of driving higher satisfaction, loyalty, retention, and business performance.

\subsection{Limitations and Future Work}

Despite its contributions, this study has several limitations that point to specific next steps. First, the final sample size (\(N = 139\)) was adequate for the primary regression and clustering analyses. However, future research should recruit larger, more heterogeneous cohorts, including cross-cultural and domain-specific populations, to enable finer-grained moderation analyses and broader generalisability.
Second, the use of scripted dialogues provided experimental control but abstracted from live CRS dynamics. To address this, our forthcoming prototype study will embed the threshold-based adaptive heuristics described above into a working conversational system. This implementation will support real-time system–user co-construction of recommendations and allow us to evaluate effects on user satisfaction and behavioural intentions (for example, continued use, purchase intent, or recommendation to others).
Finally, we examined two distinct dialogue variants within a single domain. While this scope enabled rigorous hypothesis testing, subsequent work should extend this paradigm to multiple application areas and explore additional interaction styles, such as mixed-initiative or summarisation-driven flows. This would help assess the robustness and scalability of our adaptive strategies across longer-term engagements.

\section{Conclusion}

Our study advances the field of conversational recommender systems by shifting from retrospective, \textit{post hoc} evaluation toward a predictive and adaptive framework that integrates system qualities, user experience appraisals, and individual‐difference traits. We show that novelty, conversational quality, and usefulness can predict whether a user prefers a \emph{task-oriented} or \emph{exploratory} interaction style, thereby extending the CRS-Que framework into the domain of preference modelling. Clustering analyses uncovered five latent user profiles, supporting value-based segmentation as an effective personalisation strategy in cold-start scenarios where behavioural histories are unavailable. Furthermore, moderation tests revealed that age, gender, and control preferences significantly influence how system quality beliefs shape users’ dialogue style choices.

Importantly, the results challenge the common assumption that instrumental attributes such as usefulness and effectiveness are inherently aligned with \emph{task-oriented} interactions. Instead, these attributes, when understood as markers of informational completeness and clarity, were more strongly associated with exploratory dialogues that promote sensemaking and user empowerment. This finding suggests that instrumental UX dimensions may reflect both cognitive and affective evaluations. Accordingly, the design of \emph{task-oriented} interactions should be reconsidered to include contextual support mechanisms, such as adaptive summarisation and information-density controls, that help users process and act on recommendations effectively.

From a practical perspective, our work offers design principles for the next generation of CRS interfaces powered by generative AI. By applying rule-based switching strategies informed by real-time indicators of user needs (for example, novelty sensitivity, enjoyment, or desire for control), designers can enable systems to shift fluidly between \emph{task-oriented} and \emph{exploratory} modes. Such adaptive systems are likely to improve short-term satisfaction and long-term engagement. Although this study focused on recommendation contexts, the predictive and adaptive framework proposed here is applicable to a broader class of conversational AI systems, including intelligent tutoring, customer support, and personal assistants, where tailoring dialogue to the user’s cognitive and emotional state can similarly enhance outcomes.

By contributing theoretical insights and design patterns, and by preparing for future evaluations in real-world settings, this study lays the groundwork for conversational agents that do more than recommend. These systems can learn from each user’s values, expectations, and emotional responses, and adapt accordingly, thereby delivering personalised experiences that are context-sensitive, transparent, and responsive to the complexity of human needs.

\section*{Acknowledgements}

During the preparation of this work the author(s) used ChatGPT 4o in order to improve readability and language of the final draft. After using this tool/service, the author(s) reviewed and edited the content as needed and take(s) full responsibility for the content of the published article.




\appendix
\renewcommand{\thefigure}{A.\arabic{figure}}
\setcounter{figure}{0}

\renewcommand{\thetable}{A.\arabic{table}}
\setcounter{table}{0}

\section{Appendix: Scripted Dialogue Excerpts}
\label{sec:scripted_dialogues}

This appendix provides representative excerpts from the two scripted dialogues used as stimuli in the study, illustrating the key differences between the \emph{task-oriented} and \emph{exploratory} interaction described in Section \ref{subsec:interaction-styles} and operationalised in Section \ref{sec:research-design}

\subsection{Task-oriented Interaction}

\begin{quote}
\textbf{System:} Welcome! What type of dress are you looking for today?

\textbf{User:} I want a casual dress.

\textbf{System:} Great! Here are the top 3 casual dresses we recommend. The first one is priced at \$59.99 with a 10\% discount.

\textbf{User:} I like the first one.

\textbf{System:} Excellent choice! Would you like to add it to your cart?
\end{quote}

This dialogue illustrates the concise and goal‐directed nature of the \emph{task-oriented} interaction, where the system quickly narrows down options and facilitates rapid decision‐making with minimal elaboration.

\subsection{Exploratory Interaction}

\begin{quote}
\textbf{System:} Hello! Looking for a dress today? What styles do you usually enjoy?

\textbf{User:} I like casual dresses mostly.

\textbf{System:} That’s wonderful! Casual dresses can range widely from flowy maxi dresses to chic shirt dresses. Are you interested in exploring some trendy new arrivals or classic styles?

\textbf{User:} Maybe something trendy.

\textbf{System:} Perfect! How about a light floral maxi dress? It’s popular this season and perfect for both casual outings and special occasions. I can also suggest some accessories to complement the look.

\textbf{User:} Sounds interesting! Show me more options.

\textbf{System:} Certainly! Here are some other trendy choices along with styling tips.
\end{quote}

This excerpt demonstrates the \emph{exploratory} interaction’s open‐ended prompts, elaborative responses, and reflective dialogue, designed to encourage discovery and deliberation.

\clearpage
\section{Appendix: Clustering Diagnostics}
\label{clustering-elbow-silhouette}

\begin{figure}[h]
  \centering
  \begin{tabular}{cc}
    \includegraphics[width=0.42\linewidth]{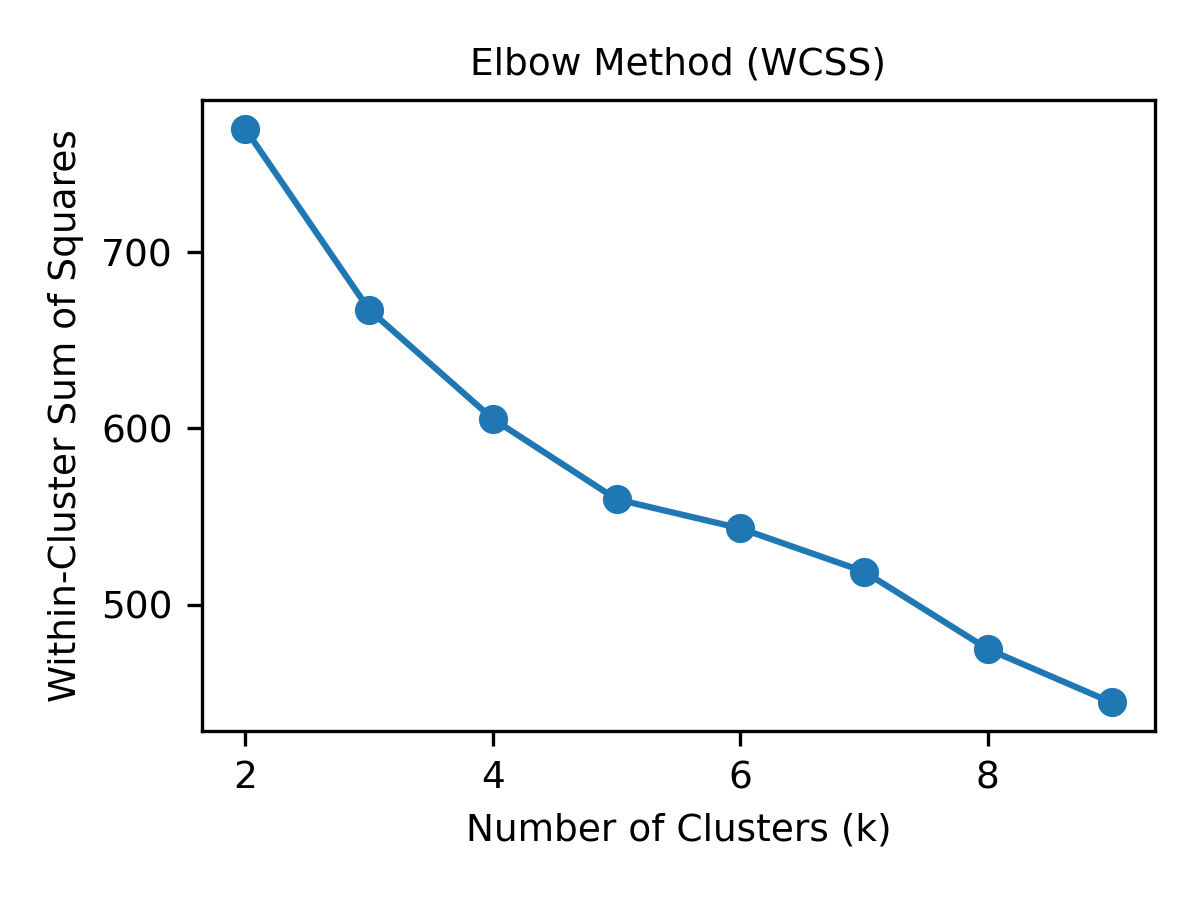} &
    \includegraphics[width=0.42\linewidth]{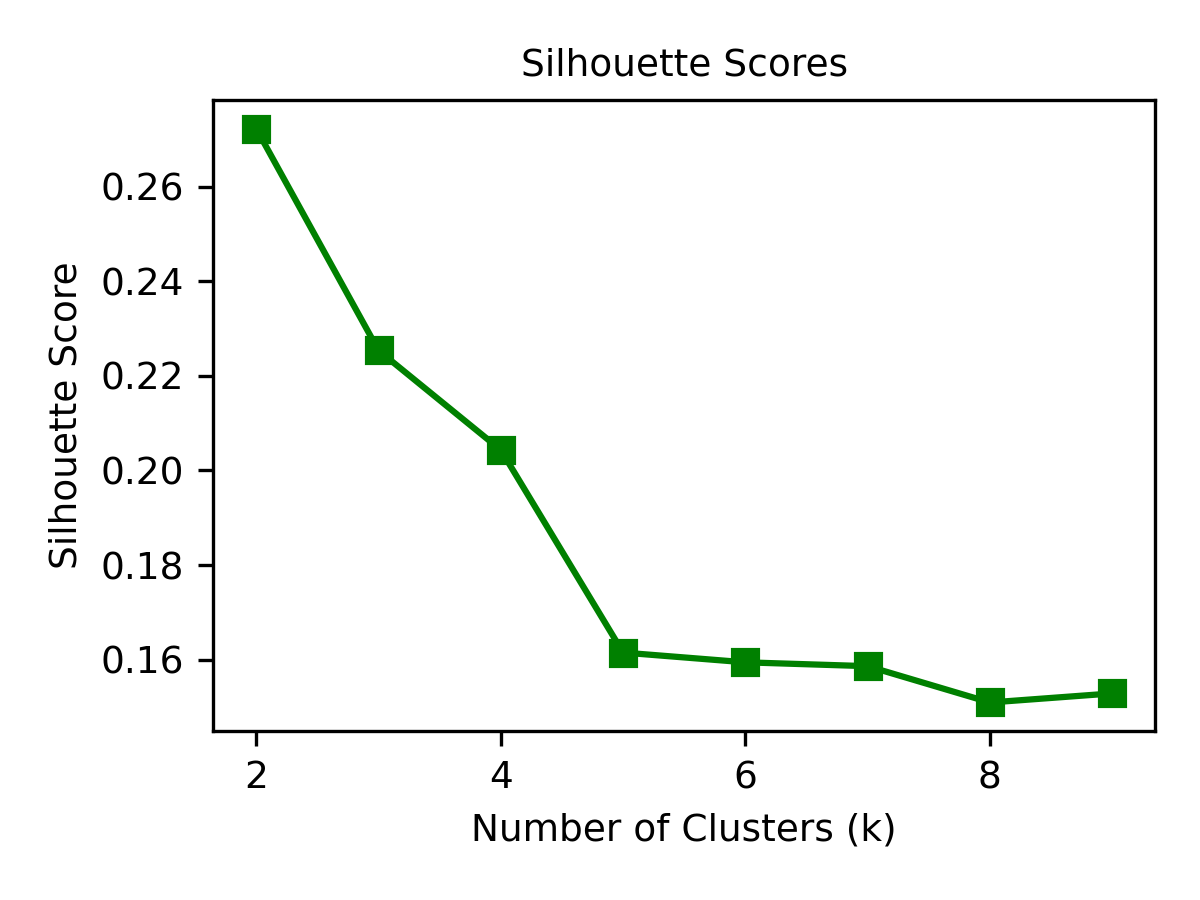} \\
    (a) Elbow plot & (b) Silhouette scores \\
  \end{tabular}
  \caption{
    Clustering diagnostics supporting the five‐cluster solution for RQ2. (a) Elbow plot of within‐cluster sum of squares (WCSS) for \(k = 2\) through \(9\). The inflection at \(k=5\) indicates diminishing returns in variance reduction beyond five clusters. (b) Silhouette scores for \(k = 2\) through \(9\), which plateau near \(k=5\), indicating that five clusters offer a reasonable compromise between cohesion and separation.}
  \label{fig:appendix_rq2_clustering_diagnostics}
\end{figure}

In Section~\ref{sec:results-cluster-profiles}, we demonstrated that a five-cluster solution most effectively differentiates participants by their CRS-Que importance ratings. Figure~\ref{fig:appendix_rq2_clustering_diagnostics} presents the elbow plot (within-cluster sum of squares) and silhouette scores for \(k=2\) through \(9\). These diagnostics, together with the stability and preference-association metrics in Table~\ref{tab:appendix_cluster_diagnostics}, confirm that \(k=5\) achieves an optimal compromise between cluster cohesion and separation.

\begin{table}[ht]
  \caption{Diagnostics for hierarchical (Ward’s) clustering solutions, $k=3\text{–}9$ ($N=139$). “Min/Max” denotes the smallest and largest cluster sizes. Silhouette is the average silhouette coefficient. ARI mean and SD report perturbation‐bootstrap stability. $\chi^2$ and $H$ tests assess the association between cluster membership and interaction preference.}
  \label{tab:appendix_cluster_diagnostics}
  \centering
  \small
  \begin{tabular}{r r r r r r r r r}
    \toprule
    $k$ & Min/Max & Silhouette & ARI mean & ARI sd & $\chi^2$ & $p_{\chi^2}$ & $H$ & $p_H$ \\
    \midrule
    3 & 18/87  & 0.182 & 0.730 & 0.287 & 14.92 & 0.0006 & 14.81 & 0.0006 \\
    4 & 18/63  & 0.150 & 0.747 & 0.203 & 14.93 & 0.0019 & 14.82 & 0.0020 \\
    5 & 18/37  & 0.146 & 0.770 & 0.214 & 14.93 & 0.0048 & 14.82 & 0.0051 \\
    6 & 3/37   & 0.153 & 0.714 & 0.185 & 15.34 & 0.0090 & 15.23 & 0.0094 \\
    7 & 3/34   & 0.127 & 0.731 & 0.185 & 15.38 & 0.0175 & 15.27 & 0.0182 \\
    8 & 3/34   & 0.131 & 0.774 & 0.190 & 15.38 & 0.0314 & 15.27 & 0.0327 \\
    9 & 3/26   & 0.117 & 0.777 & 0.168 & 16.03 & 0.0420 & 15.91 & 0.0436 \\
    \bottomrule
  \end{tabular}
\end{table}

\clearpage
\section{Appendix: Cluster Profiles}

\begin{figure}[h]
  \centering
  \includegraphics[width=0.8\linewidth]{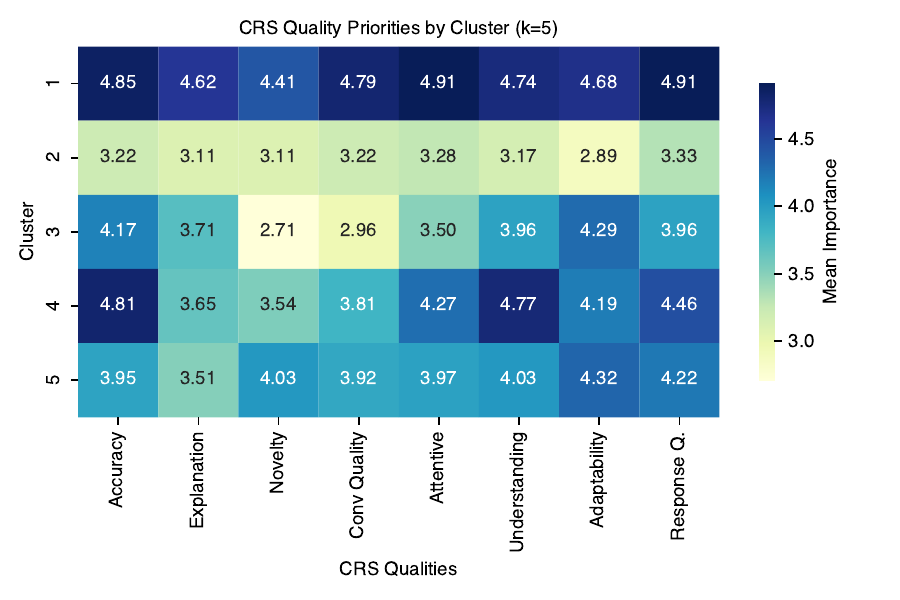}
  \caption{Heatmap of mean importance ratings (1 = low to 5 = high) for eight CRS qualities across five user clusters (hierarchical Ward’s method, $k=5$, $N=139$). Darker cells indicate higher cluster-level preference.}
  \label{fig:rq2_cluster_profiles}
\end{figure}

In Section~\ref{sec:results-cluster-profiles} of the main text, we described a hierarchical (Ward’s method) clustering of $N=139$ participants based on their z-standardised importance ratings for eight CRS-Que qualities. Figure~\ref{fig:rq2_cluster_profiles} visualises the resulting five clusters: each cell shows the mean rating for a given quality—accuracy, explainability, novelty, conversational quality, attentiveness, understanding, adaptability, and response quality—within that cluster.

\end{document}